# A low density of 0.8 g cm$^{-3}$ for the Trojan binary asteroid 617 Patroclus

**Franck Marchis[1], Daniel Hestroffer[2], Pascal Descamps[2], Jérôme Berthier[2], Antonin H. Bouchez[3], Randall D. Campbell[3], Jason C. Y. Chin[3], Marcos A. van Dam[3], Scott K. Hartman[3], Erik M. Johansson[3], Robert E. Lafon[3], David Le Mignant[3], Imke de Pater[1], Paul J. Stomski[3], Doug M. Summers[3], Frederic Vachier[2], Peter L. Wizinovich[3] & Michael H. Wong[1]**

[1]*Department of Astronomy, University of California, 601 Campbell Hall, Berkeley, California 94720, USA.* [2]*Institut de Mécanique Céleste et de Calculs des Éphémérides, UMR CNRS 8028, Observatoire de Paris, 77 Avenue Denfert-Rochereau, F-75014 Paris, France.* [3]*W. M. Keck Observatory, 65-1120 Mamalahoa Highway, Kamuela, Hawaii 96743, USA.*

**The Trojan population consists of two swarms of asteroids following the same orbit as Jupiter and located at the L4 and L5 stable Lagrange points of the Jupiter–Sun system (leading and following Jupiter by 60°). The asteroid 617 Patroclus is the only known binary Trojan[1]. The orbit of this double system was hitherto unknown. Here we report that the components, separated by 680 km, move around the system's centre of mass, describing a roughly circular orbit. Using this orbital information, combined with thermal measurements to estimate the size of the components, we derive a very low density of $0.8^{+0.2}_{-0.1}$ g cm$^{-3}$. The components of 617 Patroclus are therefore very porous or composed mostly of water ice, suggesting that they could have been formed in the outer part of the Solar System[2].**

The existence of binary asteroidal systems has been confirmed observationally during the past decade with spacecraft exploration[3], ground-based imaging[4], radar observations[4], and light curve measurements[5]. Most recently, the discovery of two moons orbiting around the irregular rubble pile asteroid 87 Sylvia[6] with adaptive optics system observations confirms that collision and disruption is the main formation mechanism for multiple main-belt systems. The system 617 Patroclus, the only binary Trojan known[1], was discovered with the Hokupa'a Gemini 8-m adaptive optics system[7] under excellent seeing conditions. Because of their faintness (with a magnitude in the visible spectrum of $m_v$>15.5), Trojan asteroids cannot be directly observed by most of the adaptive optics systems. Using related techniques such as stellar appulse[8] and Laser Guide Star (LGS) observations[9] at the Lick 3-m telescope, our group failed to detect any new companions of Trojans, indicating that the proportion of multiple systems in the Trojan population larger than 40 km in diameter is less than 4% (ref. 9). The study of a companion's orbit provides information about the mass, density and porosity of a system, illuminating the role of these small bodies in the formation of our Solar System.

In 2004, a LGS[10] adaptive optics system was offered on the Keck II, a 10-m telescope at the summit of Mauna Kea in Hawaii, USA. Thanks to this technology, adaptive optics sky coverage has been expanded and a Trojan asteroid, such as 617 Patroclus (mag$_v$~16) can be observed at any time under correct seeing conditions and visibility. We initiated an observing campaign between November 2004 and May 2005, recording direct images of the double system in H (1.6 μm) and Kp (2.2 μm) broadband filters with the NIRC2 near-infrared camera (Fig. 1). Both components of the system were detected at five different epochs with an angular separation of between 45 and 190 milli-arcsecond (mas) and a difference in magnitude of Δmag≈0.17. Supplementary Table 1 summarizes the relative positions and brightnesses of the components. To refine our model and check its consistency, we included four additional observations from the Gemini Science archive taken in 2001[1] and 2002 with Hokupa'a[7] mounted on Gemini North. In excellent seeing conditions, this low-order adaptive optics system provides an angular resolution of only 85 mas.

The orbital parameters of the system were estimated independently using two algorithms for visual binaries, a Monte Carlo technique[11] (Fig. 2) and a generalized least-squares method[12]. Both methods have been successfully applied to several main-belt binary systems[13], such as 121 Hermione[14] and the moonlets of 87 Sylvia[6]. The solution with the orbital elements indicated in Table 1 is quite accurate considering 2004–2005 data with a root-mean-square residual error of 9 mas, corresponding to ~35 km. It is slightly degraded when the lower-angular-resolution 2001–2002 data provided by our colleagues[1] and from the Gemini North telescope archive are included (17 mas).

The two components, separated by 680±20 km, revolve around their centre of mass in 4.283±0.004 days in a roughly circular orbit ($e$≈0.02±0.02). Because we could obtain a purely keplerian solution matching four years of observations, we can deduce that the pole orientation of the orbit seems not to precess, and is thus probably aligned with the angular momentum pole. The low eccentricity of the orbit, compared for instance with the extremely high eccentricity (~0.8) of the binary Kuiper Belt Object[15] 1998WW31, indicates that dissipation effects, such as tides, must be considered to circularize the orbit of this double Trojan system. The angular resolution provided by the Keck LGS adaptive optics data (0.06 arcsec) is above the angular size of the components. Their size can be estimated only using near-simultaneous mid-infrared and visible observations. When the 617 Patroclus system was observed in November 2000[16], our orbital model shows that both components were clearly separated, meaning that the estimated radiometric radius ($R$) corresponds to the sum of the radii of the components through the relation $R^2 = R_1^2 + R_2^2$. Assuming the same albedo for the components (a realistic assumption because the brightness ratio in H and K bands are constant and the objects are spherical) and a size ratio of $R_1/R_2 = 1.082$, we derive a radius of $R_1 = 60.9$ km and $R_2 = 56.3$ km (with an error of 1.6 km and an albedo in the visible spectrum of $A_v = 0.04$) and a beaming parameter $\eta = 0.94$. Recent Spitzer spectra taken between 5 and 40 μm of Trojan asteroids[17] confirm that the effect of the beaming factor is negligible ($\eta \approx 1$).

Using the total mass ($M = 1.36 \times 10^{18}$ kg) derived from our orbital parameters and the size of the components estimated by radiometric measurements, we can derive for the first time the average bulk density of a Trojan P-type asteroid, a class of object whose interior composition is generally not well known but may possibly include organic-rich silicates, carbon and anhydrous silicates, and water ice. Even considering the uncertainty in the volume of the components which accounts for most of the error bar, the density of 617 Patroclus ($\rho = 0.8^{+0.2}_{-0.1}$ g cm$^{-3}$) is extremely low, if compared with the bulk-density of known binary C-type main-belt asteroids[13] with $\rho \approx 1.2$ g cm$^{-3}$ such as 45 Eugenia[18,19], 90 Antiope[19,20], 87 Sylvia[6], or 121 Hermione[14]. The knowledge of the mean density places a strong constraint on the internal structure and composition of this Trojan. The Trojan swarms are currently at a distance from the Sun similar to that of the Galilean satellites, so they may be made of the same material (a mixture of rock and ice such as in Ganymede and Callisto) with an uncompressed sample density[21] of ~1.6 g cm$^{-3}$. This yields a macro-porosity of ~50% (see Fig. 3), typical of loose-rubble-pile main-belt asteroids[22]. A more realistic smaller porosity of ~15% cannot be excluded by considering a different composition, such as more water ice, in the interior of 617 Patroclus. Dynamical simulations[2] show that Trojans could have formed in more distant regions and then been captured into co-orbital motion with Jupiter during the time when the giant planets migrated by interaction with a surrounding disk of planetesimals. Quantitatively, considering the observed size distribution of the Trojan population and the present bulk-density measurement of 617 Patroclus, the total mass of the Trojans calculated ($7 \times 10^{-6} M_{Earth}$) is also into the range of the total mass of their simulation[2] ($4 \times 10^{-6} M_{Earth}$ to $3 \times 10^{-5} M_{Earth}$).

An additional feature that cannot be determined by this study is the obliquity of the components' equator with respect to the determined orbital plane. It is therefore not possible to see whether the system reached its Cassini state[23] (a resonance between spin precession and orbital precession), or if the spins are prograde or retrograde with respect to the orbit, and with none or a significant obliquity. Because no complete light curve variations have been successfully recorded, the spin periods of the components remain unknown. Photometric observations[24] suggest two non-commensurate periods, implying an unsynchronized system. Because it is still unclear whether the system is synchronous, like 90 Antiope[20,25] (the only known resolved double main-belt asteroid), a special programme of photometric observations should be undertaken to attempt to derive the spin poles and rates.

The orbital angular momentum of this binary system, which would account for more than 90% of the total angular momentum if we assume a synchronous orbit, is close to 2.4 G m$^3$ R)$^{1/2}$, which is well in excess of the limit of $J_{cr}$=0.4 for binary fission of a rotating mass of fluid[26], and is also larger than that of the other known double asteroid 90 Antiope[27]. The origin of this tightly bound (the separation corresponds to ~1/60×$R_{Hill}$), small-eccentricity and large-angular-momentum binary system remains unclear. It could be the result of a capture as proposed in the case of binary minor planets in the Kuiper belt[28,29], with subsequent circularization of the orbit by tidal effects. However, a low-velocity impact of a relatively large projectile could also have supplied enough angular momentum to produce a fission of the parent body without completely dispersing it. The relative velocities in the Trojan swarms are, however, at present several kilometres per second, making such a scenario less probable. Finally, this system shares some similarities with known binary near-Earth asteroids (having small separation, mass ratios close to unity, and circular orbits), implying that it could have also formed by tidal splitting after a low-velocity encounter with a large planet[30]. Our colleagues observed in their simulations[2] that several distant planetesimals have a close approach with Jupiter before their final capture at the Lagrange points.

**Supplementary Information** is linked to the online version of the paper at www.nature.com/nature.

**Acknowledgements** This work was supported by the National Science Foundation Science and Technology Center for Adaptive Optics and by the National Aeronautics and Space Administration (NASA) issue through the Science Mission Directorate Research and Analysis programmes. Most of the data were obtained at the W. M. Keck observatory, which is operated as a scientific partnership between the California Institute of Technology, the University of California and NASA. Additional observations were obtained at the Gemini Observatory (acquired through the Gemini Science Archive).

**Author Contributions** F.M.and the IMCCE researchers processed, analysed, and interpreted the data. [NB: I move this section here again since the "Keck Science Team" are listed as authors. I rephrased it hoping to make it clearer] The 2004–2005 campaign of observations with Keck LGS adaptive optics was conducted by the team  from the W. M. Keck Observatory, and other University of California at Berkeley researchers.

**Author Information** Reprints and permissions information is available at npg.nature.com/reprintsandpermissions. The authors declare no competing financial interests. Correspondence and requests for materials should be addressed to F.M. (e-mail: fmarchis@berkeley.edu).

**Table 1 Orbital elements of the binary system 617 Patroclus**

| | |
|---|---|
| Period (days) | 4.283±0.004 |
| Semi-major axis (km) | 680±20 |
| Eccentricity, *e* | 0.02±0.02 |
| Pole solution ECJ2000 ($\lambda$ and $\beta$ in degrees) | $\lambda$=236±5 $\beta$=−61±1 |
| Pericentre argument (degrees) | 139 |
| Time of passage (Julian days, JD) | 2453171.1±0.1 |
| Total mass (kg) | $(1.36 \times 10^{18})$±0.11 |
| Bulk density of the system* (g cm$^{-3}$) | $0.8^{+0.2}_{-0.1}$ |

Two models[11,12] were used to estimate the orbital parameters (with 1$\sigma$ error bars) using 2004–2005 Keck LGS adaptive optics data and give a consistent solution. The addition of 2001 and 2002 observations taken with the Gemini North 8-m telescope and the Hokupa'a using our general least-squares method[12] improves by a factor of ten the accuracy on the period, removes the ambiguity on the pole solution of the orbit, which is retrograde, and confirms the absence of measurable precession, suggesting a low inclination of the system with respect to its invariant plane.

*Considering the radiometric measurements[16] with a beaming factor $\eta$=0.89–0.96 and assuming the shape of the two components to be spherical with $R_1$=60.9 km and $R_2$=56.3 km.

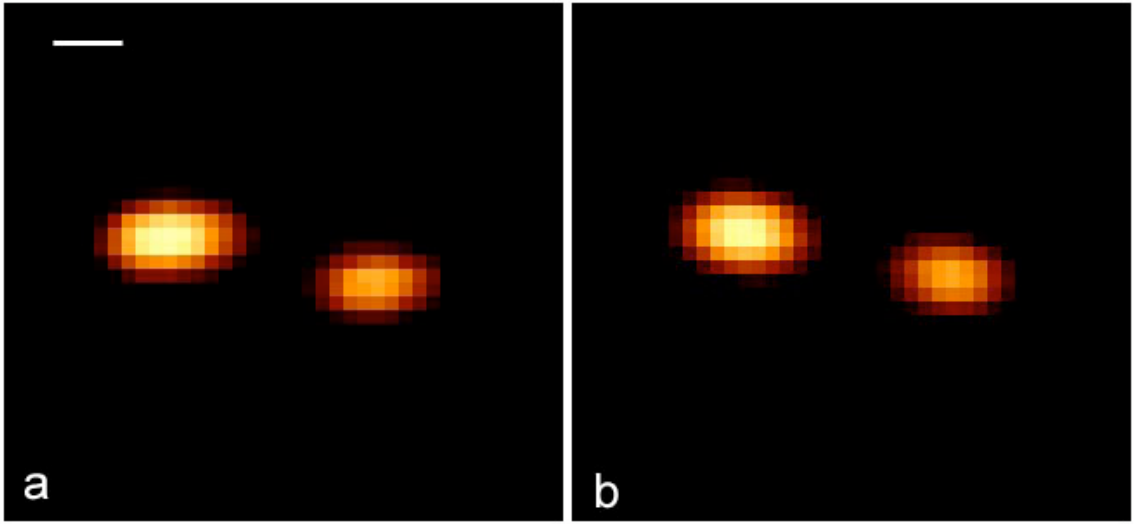

**Figure 1 617 Patroclus observed with the Keck 10-m telescope on the Mauna Kea summit in Hawaii and its adaptive optics system.** The observations of this faint object (integrated magnitude in the visible, mag$_v$≈15.8) were made possible using the LGS system. An artificial star (with $m_v$≈12) is created at proximity of the target, exciting the thin sodium layer at an altitude of ~100 km using a dye laser with a power output of ~13 W. The asteroid itself was used for the tip-tilt analysis, which is acquired by an avalanche photodiode tip-tilt sensor, and then corrected by balancing a flat mirror. The lowest-order aberrations are corrected by controlling a 384-actuator deformable mirror at a rate of 400 Hz. These observations were taken on 28 May 2005 with the NIRC-2 infrared camera through the H, centred at 1.6 μm (**a**), and Kp, centred at 2.2 μm (**b**), broadband filters. The angular resolution of the data is estimated to 58 mas, close to the diffraction limit of the telescope (indicated by the scale bar, 50 mas). In these two images, the angular separation of both components is 150 mas, corresponding to 640 km, close to the maximum separation of the system. The magnitude difference in brightness is 0.2 in both filters, suggesting a similar surface composition.

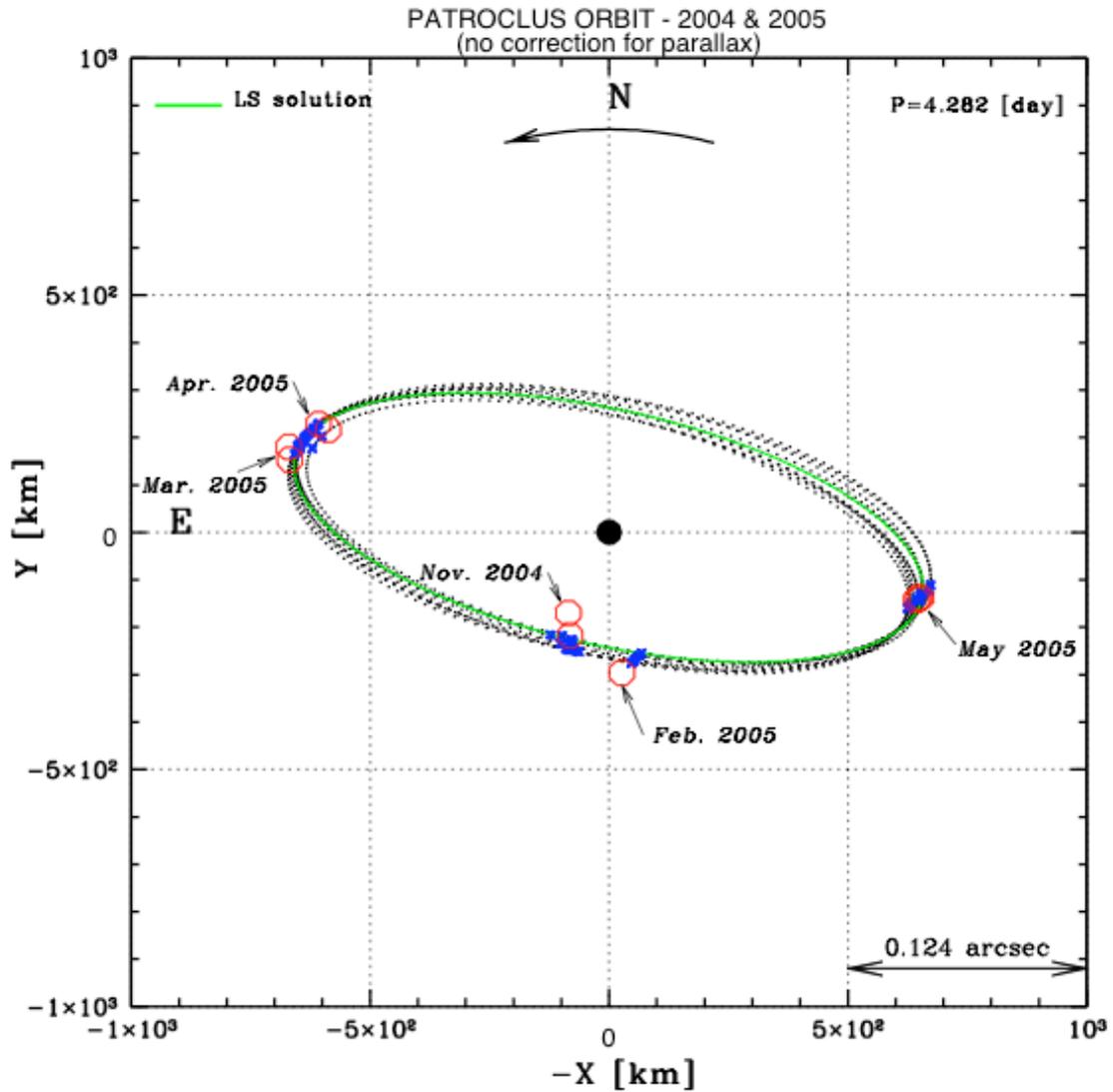

**Figure 2 Several good-fit orbits of the 617 Patroclus components estimated using 2004–2005 Keck LGS adaptive optics data.** No correction for parallax was made. The LS (Least Squareorbit (green) corresponds to the best-fit solution that minimizes the root-mean-square residual. The orbital parameters of the orbit were validated and refined by adding the Gemini/Hokupa'a observations taken in 2001 and 2002. The period is 4.283 days.

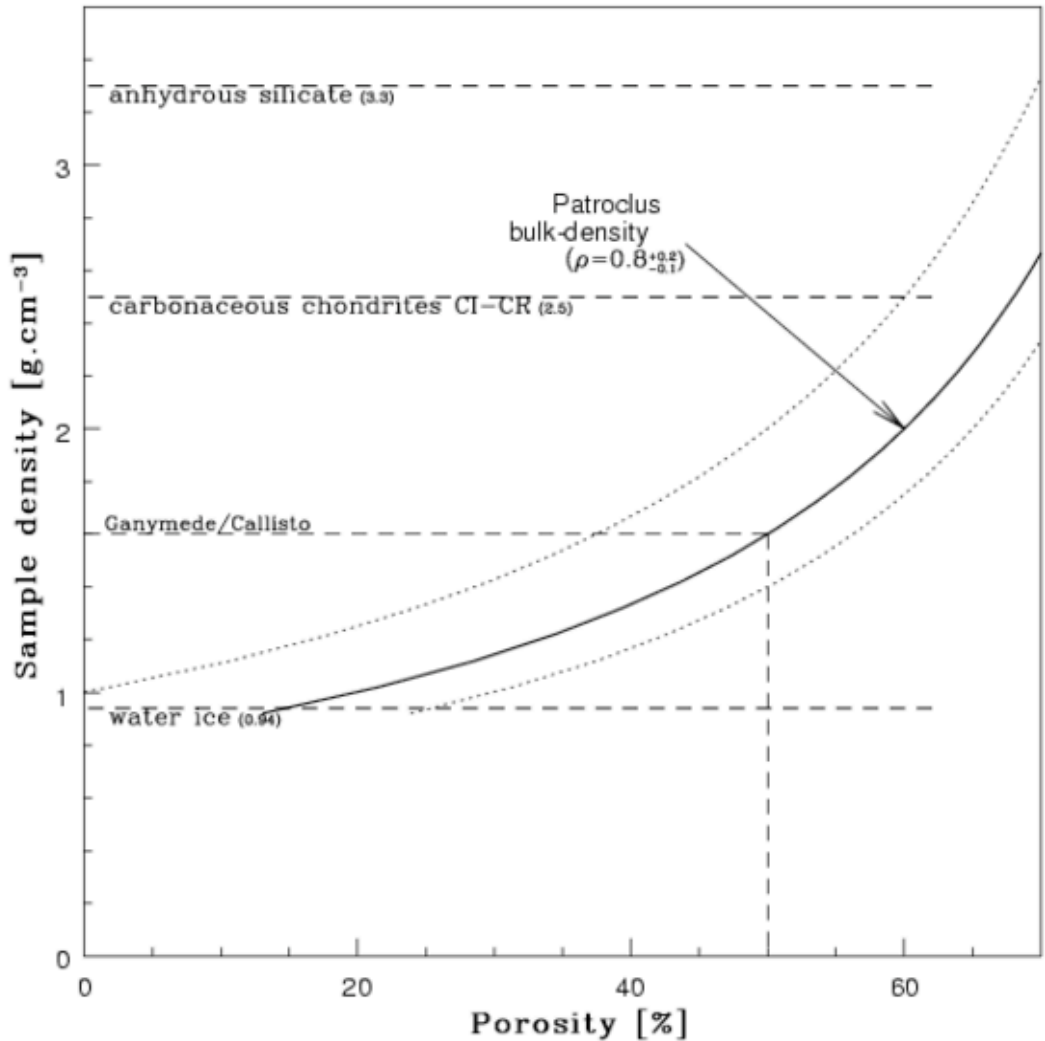

**Figure 3 Relationship between the sample density and the porosity of**

**617 Patroclus for various compositions.** Using the orbital information of the two components, combined with the thermal measurements[16], we derived a very low bulk density of ~0.8 g cm$^{-3}$ (solid line). The dotted lines correspond to the uncertainty in the bulk density for a plausible range of beaming factors varying from 0.89 to 0.96,

compatible with Spitzer Trojan spectra[17]. If 617 Patroclus components were made of the

same ice/rock mixture as Ganymede or Callisto[21], it would be extremely porous

(porosity≈50%). If they were composed of pure water ice, its porosity would be more

realistic (porosity≈15%) but would imply its formation in a more distant part of the

Solar System[2].